\documentclass[prd,tightenlines,nofootinbib,amsfonts,amssymb,amsmath,color,11pt]{revtex4}
\usepackage{graphicx}
\usepackage{color}

\usepackage[utf8]{inputenc}

\usepackage{amsfonts}
\usepackage{amsmath}
\usepackage{amssymb}
\usepackage{hyperref}

\usepackage{cancel}

\def\be{\begin{equation}} 
\def\ee{\end{equation}}
\def\bea{\begin{eqnarray}}
\def\eea{\end{eqnarray}}
\def\nn{\nonumber}

\begin{document}

\title{Runaway Quintessence, Out of the Swampland}
\author{Yessenia Olgu\'in-Trejo$^{a,c}$, Susha L. Parameswaran$^b$, Gianmassimo Tasinato$^c$, Ivonne Zavala$^c$}

\affiliation{$^a$ Instituto de F\'isica, Universidad Nacional Aut\'onoma de M\'exico,
POB 20-364, Cd.Mx.  01000, M\'exico\\
$^b$ Department of Mathematical Sciences, University of Liverpool,
Liverpool, L69 7ZL, UK \\
$^c$ Department of Physics, Swansea University, Swansea, SA2 8PP, UK}


\begin{abstract}
We propose a simple, well-motivated and robust alternative to a metastable de Sitter vacuum in string theory, consistent with current observations of dark energy and naturally satisfying conjectured swampland constraints.  Inflation ends in a supersymmetric Minkowski minimum, with a flat direction that is protected by non-renormalisation theorems.  At some scale non-perturbative effects kick in, inducing a runaway scalar potential.  The tail of this runaway potential cannot sustain a late-time dominating, slow-roll quintessence.  However, the potential always contains a dS maximum.  If the modulus starts close to the hilltop, it remains frozen there by Hubble friction for much of the cosmological history,  at first sourcing a tiny classical vacuum energy and then constituting a rolling quintessence, with observable consequences.  So long as the modulus is localised away from the Standard Model in the extra dimensions, there are no observable fifth forces nor is there time-variation of fundamental constants, and the modulus mass is protected from radiative corrections.  
We revisit concrete string models based on heterotic orbifold compactifications, and show that their de Sitter extrema satisfy the recently refined dS Swampland Conjecture.

\end{abstract}

 \maketitle


\section{Introduction}
The surprising discovery, twenty years ago, that our Universe's expansion is currently accelerating \cite{Perlmutter:1997zf,Riess:1998cb} presents arguably the most challenging problem in fundamental physics.  

There have been two main proposals to explain the phenomenon. The first is that the geometry of spacetime is de Sitter, with a tiny vacuum energy whose fine-tuning may be justified via the anthropic principle; in fact the latter led to the {\it prediction} of dark energy in \cite{Weinberg:1988cp}.  With the development of the string theory landscape (see \cite{Bousso:2012dk} for a review), this picture has found a well-motivated theoretical framework and there has been significant progress towards constructing robust examples of metastable de Sitter vacua in string theory (for an up to date review see \cite{Cicoli:2018kdo}).  However, there is not yet any metastable de Sitter vacuum under complete calculational control, and some have even conjectured that there might be no metastable de Sitter vacua in a consistent theory of quantum gravity \cite{Danielsson:2018ztv, Obied:2018sgi}.  

The second proposal is that the acceleration be driven by a slowly rolling scalar field, known as quintessence (for a review see \cite{Tsujikawa:2013fta}).  Aside from the fine-tuning required to protect both the energy density and mass of the quintessence field, quintessence models have to contend with stringent constraints on fifth forces and time-variation of fundamental constants (see e.g. \cite{Witten:2000zk}).  The main way to do so, is to assume that the quintessence field is a pseudoscalar axion. However, to source the acceleration the axion needs to have a super-Planckian axion decay constant, which has proven difficult to find within string theory \cite{svrcek} and  has been conjectured to be impossible within quantum gravity \cite{Banks:2003sx, ArkaniHamed:2006dz}.  Other string models suffer from control issues, similar to those of de Sitter (for a discussion see \cite{Akrami:2018ylq}, for a recent supergravity discussion see \cite{Chiang:2018jdg}).

In this note we offer a simple, well-motivated and robust realisation of quintessence in string theory.  Many of the supersymmetric flat directions typically encountered in string compactifcations, once lifted by leading order non-perturbative effects, experience a scalar potential for their saxion scalar component with de Sitter maximum, followed by a runaway towards decompactification or decoupling.  This leading scalar potential is protected by the classic non-renormalisation theorems \cite{Witten:1985bz, Dine:1986vd}.  Its scale may easily be multiply exponentially suppressed in the vevs of supersymmetrically stabilised moduli.  As we will show, the tail of the runaway potential cannot sustain a late-time dominating slow-roll quintessence.  However, if the initial field value is close to the de Sitter maximum, so that $(V'(\varphi_{init}))^2/V(\varphi_{init}) \ll H_{init}^2$ (for canonically normalised $\varphi$), Hubble friction freezes the field and halts the runaway for much of the cosmological history, so that it mimics a cosmological constant that comes to dominate the Universe.  Shortly after the time when  $H^2(t)\approx (V'(\varphi_{init}))^2/V(\varphi_{init})$ (which may be in the past or future), the modulus begins to roll along its runaway potential, leading to a time-dependent quintessence, with observable consequences.  The parameters can easily be chosen to match with the observed dark energy. The dS maximum is compatible with the constraints from the refined dS Swampland Conjecture \cite{Garg:2018reu, Ooguri:2018wrx} and field displacements are sub-Planckian.  Quantum fluctuations $\Delta\varphi \sim H/2\pi$ leave the field within a viable window close to the hilltop right up to $H \sim 10^{-2} M_{pl}$.  

Time variation of fundamental constants, fifth forces and radiative corrections to the quintessence mass after supersymmetry breaking in a matter sector, can all be avoided if the quintessence couples only indirectly, via gravity, to the supersymmetry (susy) breaking sector.  This may happen, e.g. if the runaway modulus describes some local geometry distant from the visible sector in the extra dimensions of the string compactification.

The axion superpartner to the runaway scalar saxion remains massless at the leading order discussed above, but is lifted by subleading non-perturbative effects to a mass less than that of the saxion - the axion thus also contributes to dark energy.  The light axino contributes to dark radiation, whose abundance depends on its production mechanism, e.g. via decay of the lightest supersymmetric modulus.

Observational constraints require that the initial conditions in this scenario are fine-tuned, with the saxion beginning close to the hilltop of its non-perturbative potential.  Moreover, we have not discussed the cosmological constant problem, which suggests a fine-tuning between the non-perturbative saxion potential energy and the vacuum energy of the Standard Model.  However, the scenario is based on a landscape of supersymmetric Minkowski vacua, with the parameters sensitive to the vevs of stabilised moduli, which in turn depend on the topological data of the internal space and fluxes.  It is therefore natural to ask if there might be an environmental anthropic explanation of both these fine-tunings, once the vacuum energy of the matter sector is taken into account.  

The note is structured as follows.  In the next section, we  give the shape of the scalar potential for a supersymmetric flat direction that is lifted by a leading non-perturbative contribution.  In Section III we show that -- with fine-tuned initial conditions -- this hilltop scalar potential can source a frozen or thawing quintessence model, which is consistent with cosmological observations.  We also comment on the possibility that the fine-tuned initial conditions might have an environmental explanation.  The next Section IV contains a brief discussion of the axion and axino superpartners to the saxion dark energy, as well as fifth forces and radiative stability of the saxion mass.  We leave for the final Section VI a discussion on the de Sitter Swampland Conjecture.

\section{Runaway Potentials for String Moduli}

Assume some inflationary (or alternative) early Universe scenario that ends in a supersymmetric Minkowski minimum, where most of the string moduli ($\Phi^i$) are stabilised and heavy:
\be
\langle D_i W_{susy} \rangle = 0 \quad \text{and} \quad \langle W_{susy} \rangle = 0 \,.
\ee
Assume a single flat direction, $\Phi=\phi+ i \theta$, with leading K\"ahler potential\footnote{Alternatively, we may consider a localised blow up modulus, with e.g. $K = K_0-2\ln(k_1 + k_2 (\Phi + \bar{\Phi})^{3/2})$. See Footnote \ref{F:blowupV}.\label{F:blowup}}:
\be
K = -n\ln(\Phi + \bar{\Phi}) \,, \label{E:K}
\ee
for example $\Phi$ may be the overall volume modulus, for which $n=3$, or it may be another volume modulus, complex structure or dilaton, for which $n=1$.

Because we are in a susy Minkowski vacuum, the flat direction is protected from perturbative corrections in string coupling constants to all finite orders, by the non-renormalisation theorems \cite{Witten:1985bz, Dine:1986vd, Burgess:2005jx, GarciadelMoral:2017vnz}.  However, non-perturbative effects will break it at some scale, say before BBN.  Consider then the leading order non-perturbative effect, exponentially suppressed in $\Phi$ (\footnote{Here $\Phi$ and $\alpha$ are dimensionless, and $A$ has dimensions of mass squared. }): 
\be
W_{np} = A e^{-\alpha \Phi} \,.
\ee
The resulting scalar potential is:
\be
V = \frac{A^2}{2^n n} e^{-2 \alpha\phi} \phi^{-n}\left(n^2 + 4 \alpha^2 \phi^2+n\left(-3+4\,\alpha\,\phi\right)\right) \,, \label{E:pot}
\ee
so the axion remains a flat direction at this order.  Corrections to this scalar potential from perturbative corrections to the K\"ahler potential, and subleading non-perturbative corrections to the superpotential will be small provided that the string coupling constants are small. 

Note that $A$ may be hierarchically small (or hierarchically large), due to an exponential suppression (or enhancement) in other, stabilised moduli.   E.g. gaugino condensation in a hidden gauge group leads to a superpotential $W=\mu^2 \exp^{-\alpha f}$, with $\mu$ the scale at which non-perturbative effects kick in, and $\alpha$ determined from the hidden sector beta function coefficient.  The gauge kinetic function $f$ may be given at tree-level by $f_{tree}=\Phi$, but can receive one-loop threshold corrections from e.g. massive winding and KK modes \cite{Kaplunovsky:1995jw, Dixon:1990pc}.  These corrections depend on other (stabilised) moduli, leading to $f= \Phi - \sum_i c_i \ln \eta(d_i\, \Phi_i)$, where $i$ runs over particular complex structure and K\"ahler moduli, $\eta(x)$ is the Dedekind eta function, $c_i$ and $d_i$ are calculable constants and $c_i$ can be positive or negative\footnote{See \cite{PRZ} for concrete examples in heterotic orbifolds.}.

Although we do not trust the above potential for all values of $\phi$, it is instructive to consider its behaviour at small and large $\phi$.  For small $\phi$, it tends to negative infinity when $n=1$ and positive infinity when $n=3$.  For large $\phi$ it tends to zero from above in both cases.  Therefore, if $n=1$ there is a dS maximum and an inflection point where $V>0$.  The dS maximum is at:
\be
\phi_{max} = \frac{1}{\sqrt{2}\alpha}\,.
\ee
and the inflection point is at:
\be
\phi_{inflex} \approx \frac{1.29}{\alpha} \,.
\ee

\begin{figure}[t!]
\begin{center}
\includegraphics[scale=0.43]{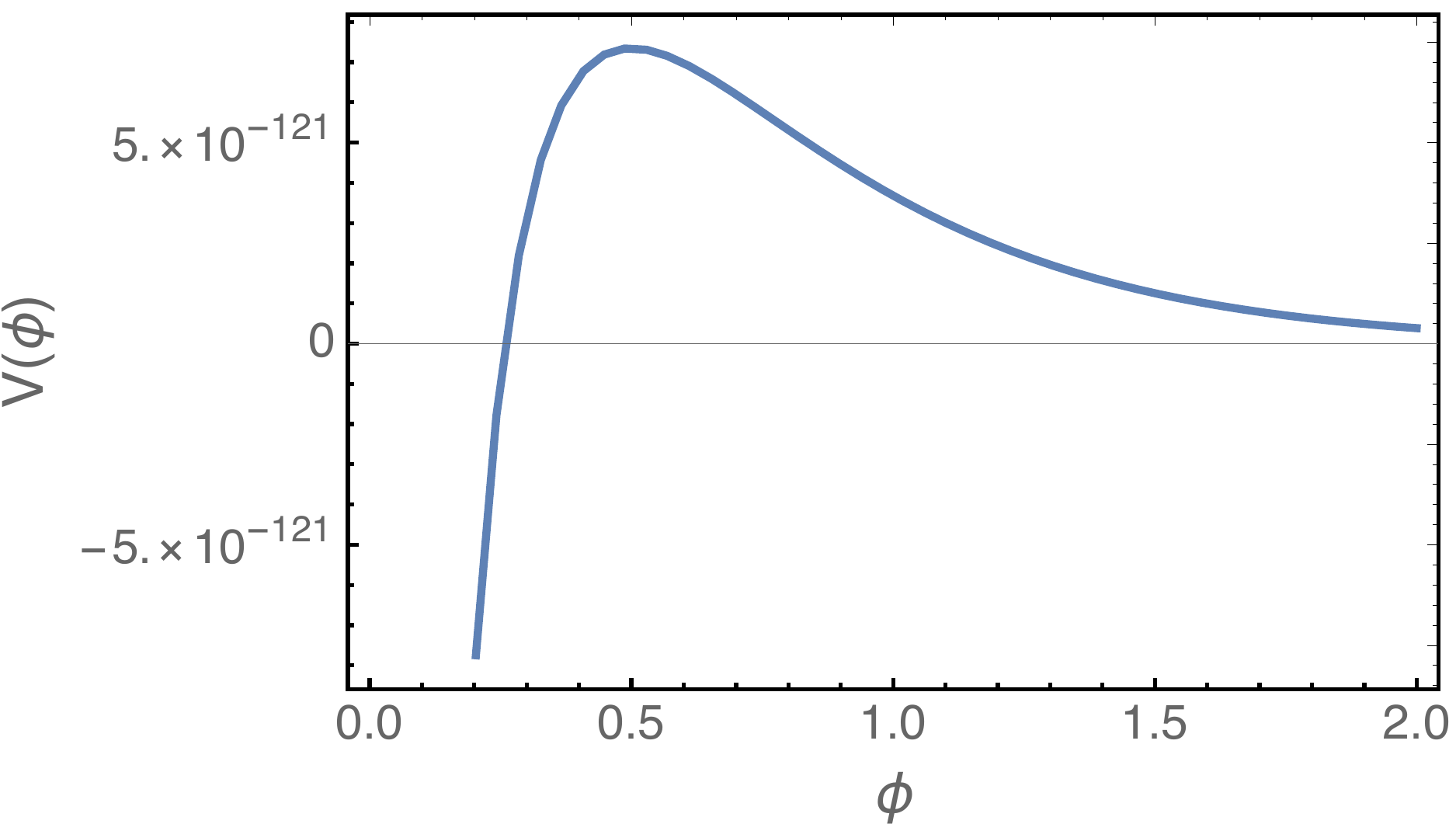}
\end{center}
\caption{Potential \eqref{E:pot} for $\alpha=\sqrt{2}$ and $A = e^{-1105/8}$ in Planck units.\label{F:pot}}
\end{figure}

From now on, our interest is in the potential $\eqref{E:pot}$ with $n=1$, or similar runaway potentials\footnote{For the blowup modulus, discussed in Footnote \ref{F:blowup}, $V \rightarrow -\frac{3 A^2 e^{K_0}}{k_1^2}$ for $\phi \rightarrow 0$ and $V \rightarrow e^{-2a\phi} \frac{a^2 A^2 e^{K_0}}{6 k_2^2 \phi}$ for $\phi \rightarrow \infty$, and therefore the potential again has a dS maximum. \label{F:blowupV}} with dS maximum. We plot the potential $\eqref{E:pot}$ in Figure \ref{F:pot}.

\section{Quintessence from a Runaway Modulus} \label{S:hilltop}
The runaway modulus is governed by the cosmological equations: 
\bea
&&3 \left(\frac{\dot{a}}{a}\right)^2 = \frac12 \frac{\dot{\phi}^2}{\phi^2} + M_{pl}^{-2}V + 3 H_0^2 \Omega_M a(t)^{-3} + 3 H_0^2 \Omega_r a(t)^{-4} \,  \nn \\
&&0 = \ddot{\phi} + 3 \frac{\dot{a}}{a} \dot{\phi} + \Gamma^\phi_{ab} \dot{\phi}^a \dot{\phi}^b + M_{pl}^{-2}g^{\phi b} \frac{\partial V}{\partial \phi^b} \,\nn\\ 
&&0 = \ddot{\theta} + 3 \frac{\dot{a}}{a} \dot{\theta} + \Gamma^\theta_{ab} \dot{\phi}^a \dot{\phi}^b + M_{pl}^{-2}g^{\theta b} \frac{\partial V}{\partial \phi^b} \, , \label{E:Frie}
\eea
where $H_0$ is today's Hubble constant, $\Omega_M$ is the density parameter for matter and $\Omega_r$ the one for radiation, with $a(t_{today})=1$.  Also, $g_{ab}$ and $\Gamma^a_{bc}$ are target-space metric and Christoffel symbols derived from the K\"ahler potential \eqref{E:K}.
 
Following these equations, under certain conditions the runaway modulus can source an accelerated expansion at late times.  In particular, the system includes a slow-roll quintessence scenario \cite{Chiba:2009sj}, where $\frac12 \dot{\varphi}^2 \ll V$ for the canonically normalised field:
\be 
\varphi = M_{pl} \sqrt{\frac{n}{2}}\log{\phi} \,.
\ee
Since the potential for the axion, $\theta$, is flat, we consider the solution $\dot{\theta}=0$. The slow-roll condition for the (non-canonically normalised) saxion field, $\phi$, is then:  
\be
\frac{M_{pl}^2}{4} \frac{\dot{\phi}^2}{\phi^2} \ll V \, \,,\label{E:SR}
\ee
whereas the Klein-Gordon equation \eqref{E:Frie} gives:
\be
\dot{\phi} \approx \frac{2\phi^2 V'(\phi) M_{pl}^{-2} -\ddot{\phi}}{3H}\,,
\ee
where $H=\dot a/a$, so (neglecting $\ddot{\phi}$) $\phi$ is slowly rolling as long as:
\be
2\phi^2 \frac{V'(\phi)^2}{V} \ll M_{pl}^{2} H^2 \,. \label{E:notail}
\ee

It is instructive then to study the behaviour of the slow-roll parameter, $2 \phi^2 V'(\phi)^2/V(\phi)$ in different regions of the potential:
\bea
2 \phi^2 \frac{V'(\phi)^2}{V(\phi)}&\rightarrow & -\frac{2 A^2}{\phi} \quad \text{as} \quad \phi \rightarrow 0\,, \\
2 \phi^2 \frac{V'(\phi)^2}{V(\phi)}&\rightarrow& 16(3+2\sqrt{2})A^2 e^{-\sqrt{2}}\alpha^3(\phi-\frac{1}{\sqrt{2}\alpha})^2 \quad \text{as} \quad \phi \rightarrow \phi_{max}\,, \\
2 \phi^2 \frac{V'(\phi)^2}{V(\phi)} &\rightarrow& 1.7 A^2 \alpha\quad \text{as} \quad \phi \rightarrow \phi_{inflex}\,,\\
2 \phi^2 \frac{V'(\phi)^2}{V(\phi)} &\rightarrow & e^{-2 \alpha \phi} 16 A^2 \alpha^4 \phi^3 \quad \text{as} \quad \phi \rightarrow \infty
\,.
\eea

Clearly, if the initial value of the modulus is near the hilltop, $\phi_{init}\approx \frac{1}{\sqrt{2}\alpha}$, it remains frozen there by Hubble friction for much of the cosmological history, sourcing a cosmological constant.  As the Hubble expansion parameter, $H$, decreases, eventually $M_{pl}^2H^2 \lesssim 2 \phi_{init}^2 V'(\phi_{init})^2/V(\phi_{init})$, and the field begins to roll, playing the role of dynamical quintessence.  See \cite{Dutta:2008qn} for an earlier study of hilltop quintessence.

One may also ask if a viable quintessence model can be found at the tail of the runaway.  For this to be possible, we need that the slow-roll parameter, $2 \phi^2 \frac{V'(\phi)^2}{V(\phi)}$ is still comparable to $M_{pl}^2 H^2$ when the dominant contribution to $M_{pl}^2 H^2$ is from $V(\phi)/3$.  In other words we need $6 \phi^2 \frac{V'(\phi)^2}{V(\phi)^2} \lesssim \mathcal{O}(1)$, whereas for the runaway modulus we have\footnote{Note that taking $\phi$ large and $\alpha$ small, such that $\alpha \phi \to 0$ does not help; in this case $6 \phi^2 \frac{V'(\phi)^2}{V(\phi)^2} \rightarrow 6$.}:
\be
6 \phi^2 \frac{V'(\phi)^2}{V(\phi)^2} \rightarrow 24 \alpha^2 \phi^2 \quad \text{as} \quad \phi \rightarrow \infty \,.
\ee
This means that -- at the tail -- if $V$ dominates the energy density of the Universe, then we cannot have slow-roll quintessence, or if there is slow-roll quintessence, then it cannot be the dominant contribution to the energy density of the Universe.  It would be interesting to investigate if there are other string theory motivated scalar potentials where late-time dominating quintessence can be sourced at the tail, perhaps even providing a quintessential inflation model \cite{Peebles:1998qn}, see however Section \ref{S:swamp}.

In summary, if the runaway string modulus has initial conditions near the hilltop, it yields a frozen or thawing quintessence model, potentially consistent with the observed late-time dark energy, characterised by a time-dependent equation of state parameter:
\be
\omega = \frac{\frac12 \dot{\varphi}^2 - V(\varphi)}{\frac12 \dot{\varphi}^2 + V(\varphi)} \,.
\ee

\begin{figure}[t!]
\begin{center}
\includegraphics[scale=0.30]{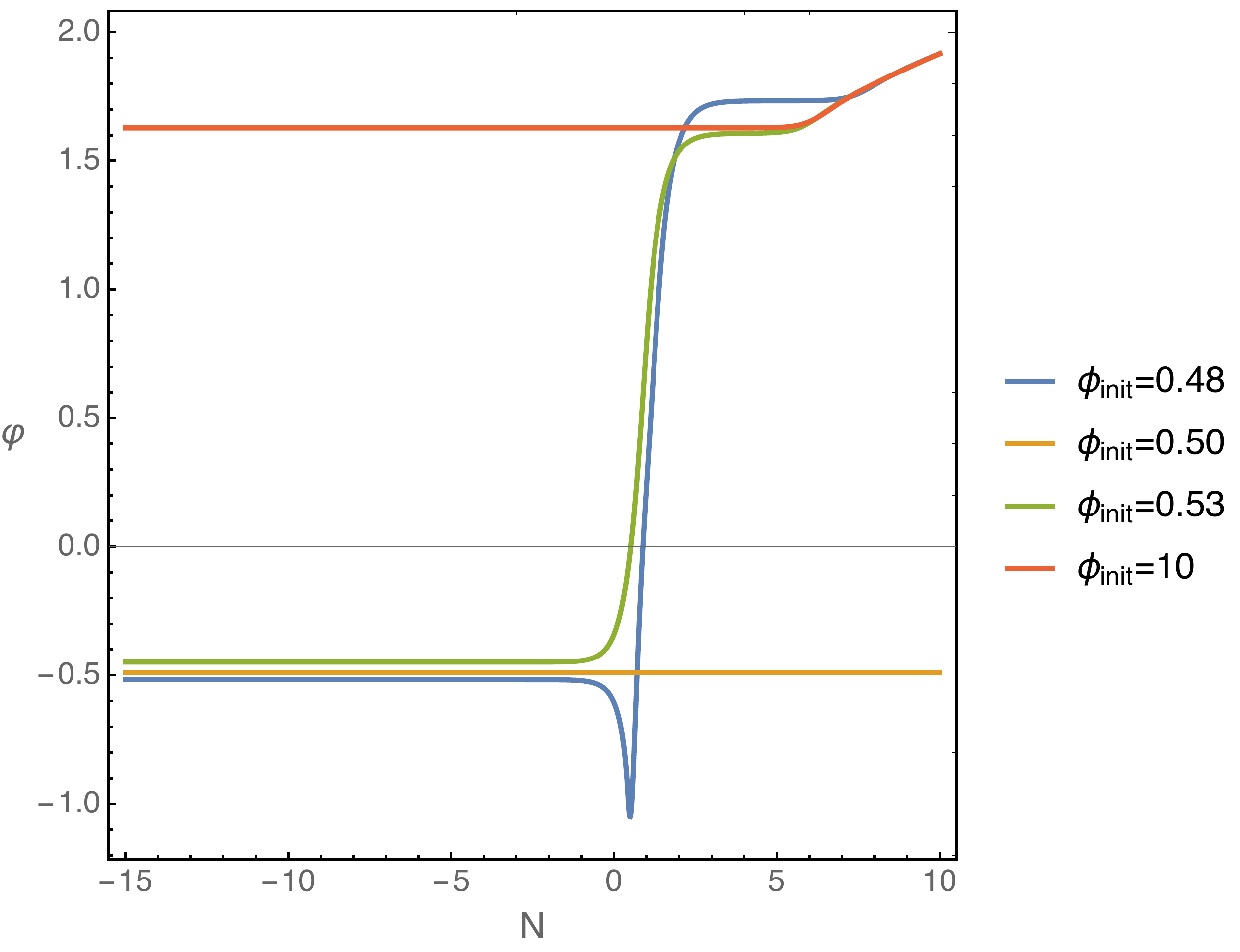}\includegraphics[scale=0.33]{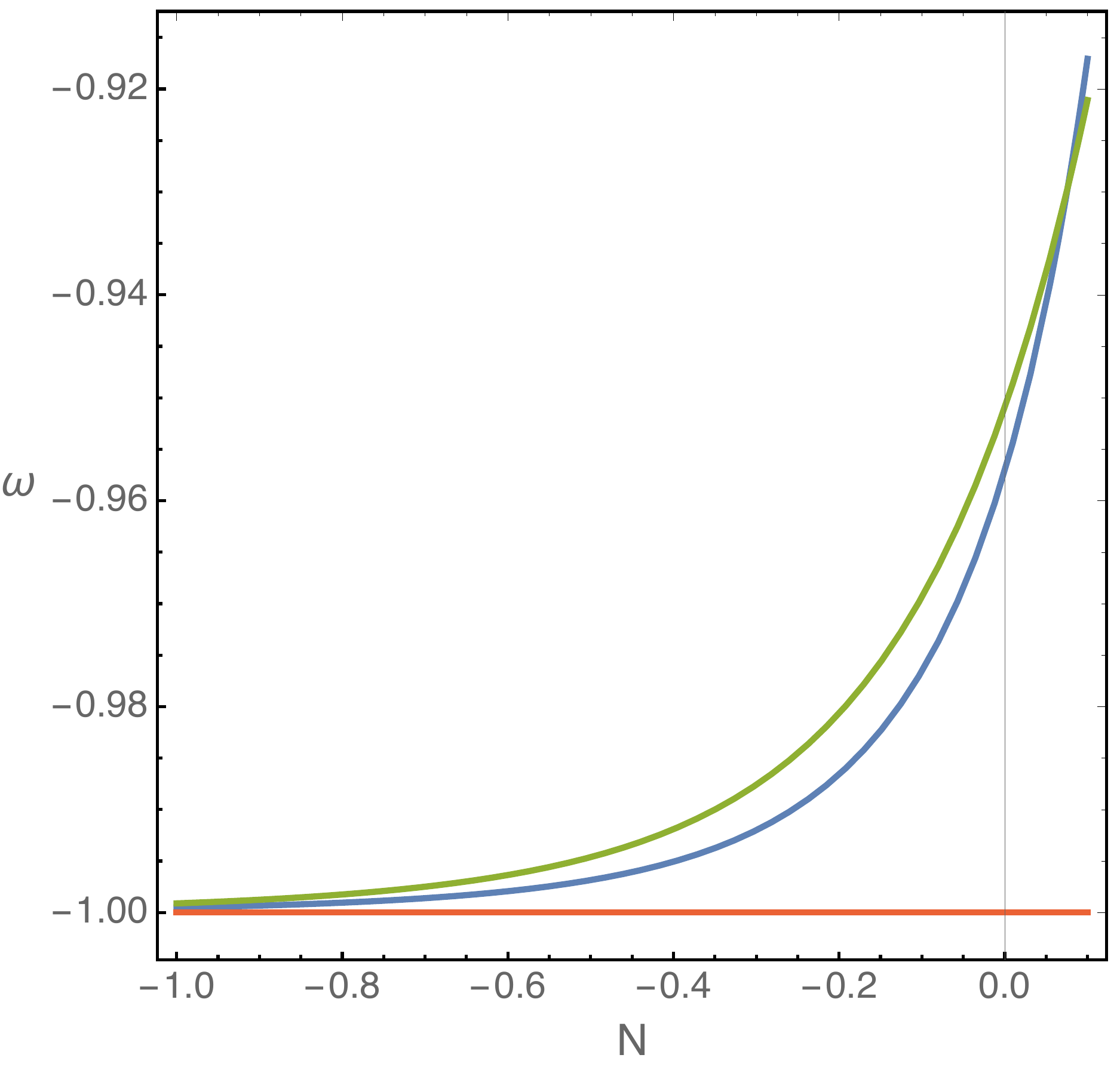} \qquad
\includegraphics[scale=0.33]{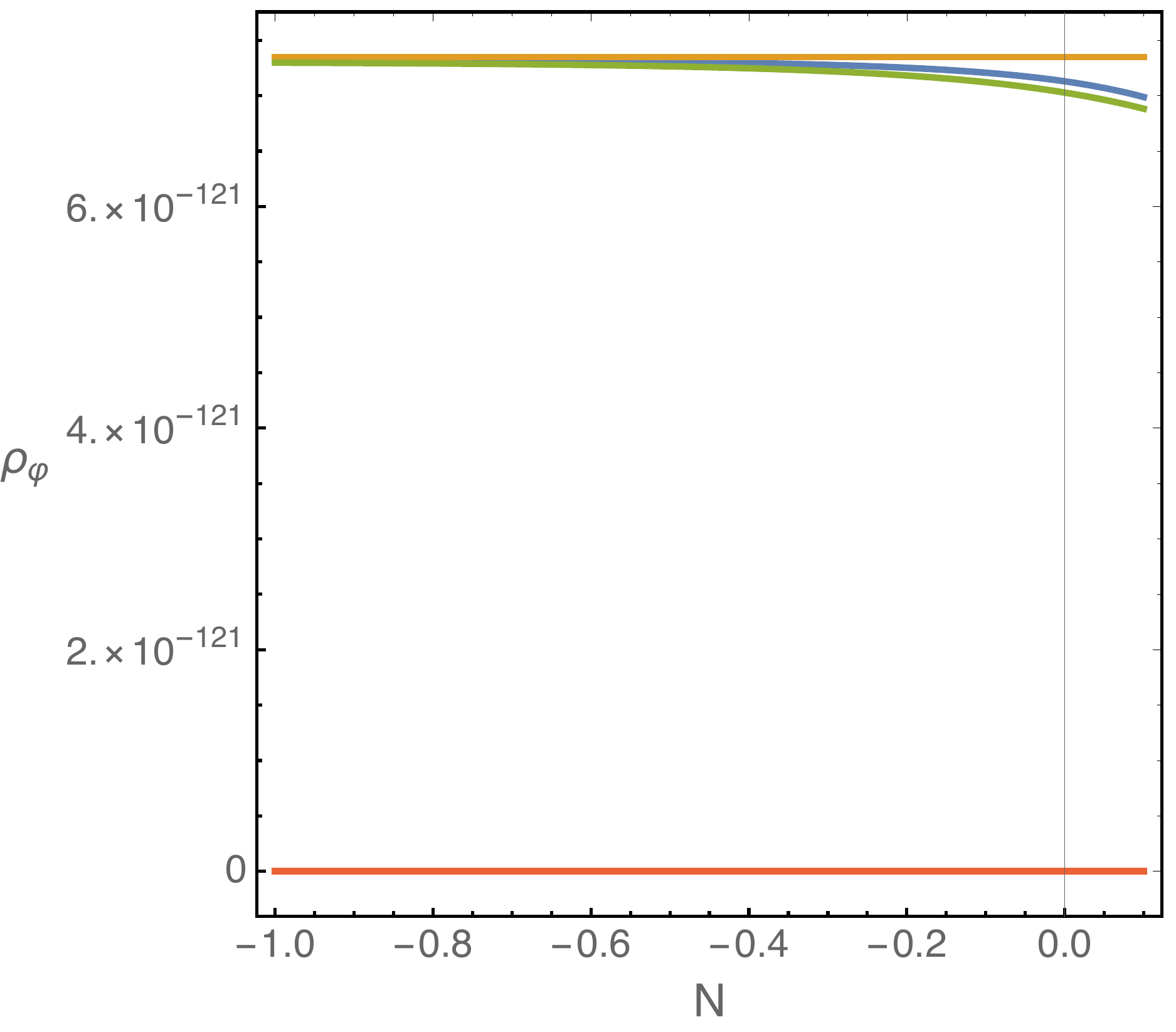}\qquad \qquad\includegraphics[scale=0.33]{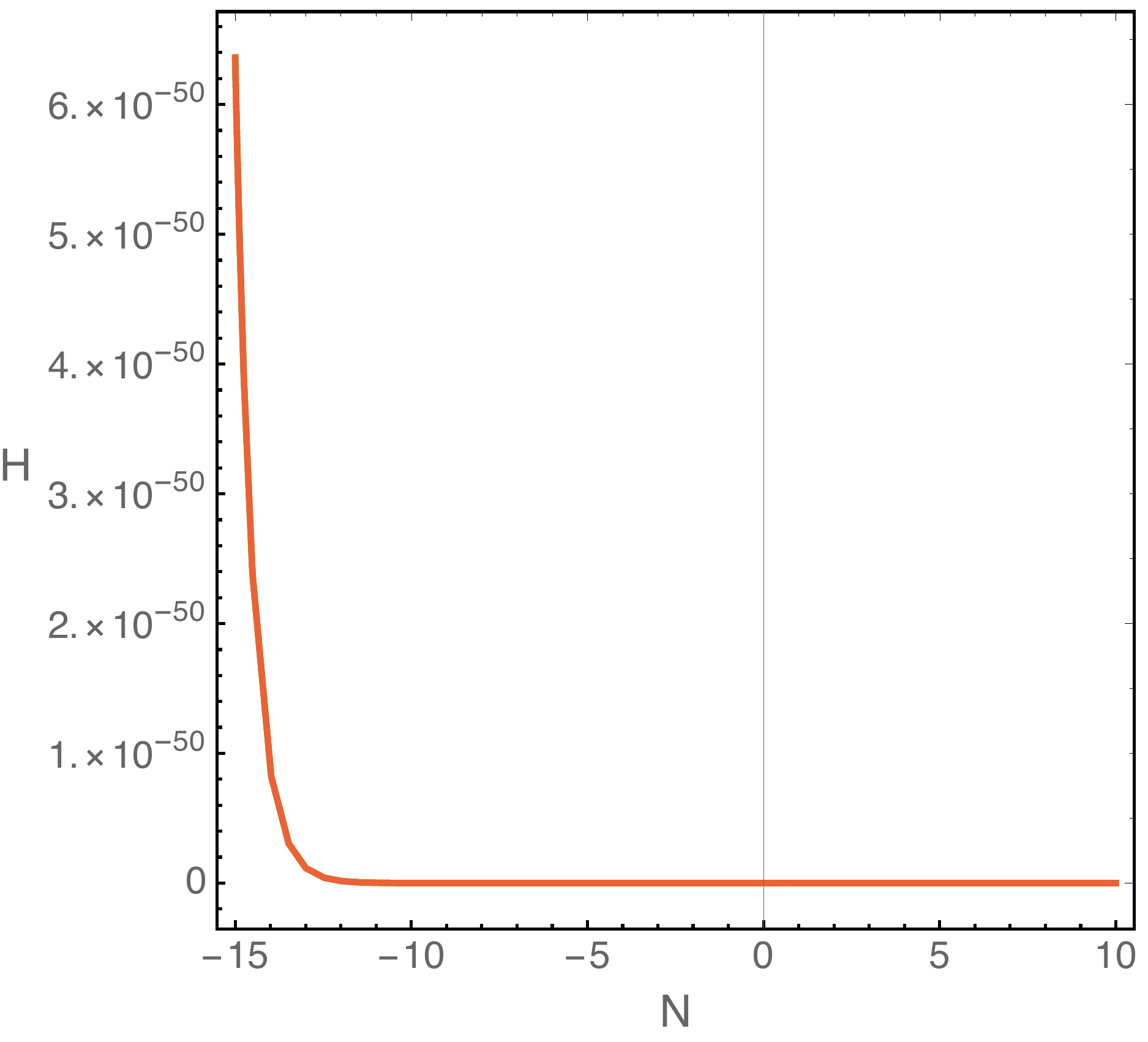}
\end{center}
\caption{Solution to the cosmological equations \eqref{E:Frie} with potential \eqref{E:pot} and parameters $\alpha=\sqrt{2}$ and $A = e^{-138.122}$, choosing various initial conditions $\phi_{init}$, $\dot{\phi}_{init}=0$ and $\dot{\theta}_{init}=0$.  We take $H_0=5.95 \times 10^{-61} M_{pl}$, $\Omega_M=0.31$ and $\Omega_r = 10^{-4}$.  We plot variables with respect to the number of e-folds, with $N=0$ today, and $N_{init}=-15$ is some time between BBN and matter-radiation equality.\label{F:thawing}}
\end{figure} 

In Figure \ref{F:thawing}, we choose parameter $\alpha = \sqrt{2}$, and plot the evolution\footnote{It is convenient here to solve the equations of motion, \eqref{E:Frie}, by rewriting them using the independent variable $N$.} of $\varphi$, $\omega$ and $\rho_\varphi = \frac12 \dot{\varphi}^2 + V(\varphi)$ with respect to the number of e-folds, $N=\ln a$.  With initial condition, $\dot{\theta}_{init}=0$, the axion, $\theta$, remains frozen at its initial value. For an initial value at the hilltop, $\phi_{init} = \phi_{max} = 0.5$, the saxion remains frozen there and the model is indistinguishable from a cosmological constant, with $\omega=-1$.  The parameter $A$ can be determined by matching to the observed dark energy scale, $A = \exp^{-138.122}$, giving today $\rho_\varphi=7.35 \times 10^{-121}M_{pl}^4$ and $H = 1.45 \times 10^{-33}eV$.  The masses for the canonically normalised fields are given by $m^2=0$ and $-\left( 5.45 \times 10^{-33}eV\right)^2$.  For $\phi_{init}$ to within around 4\% of the hilltop value $\phi_{max} = 0.5$, the evolution is consistent with current observations.  For example, for $\phi_{init}=0.48$, we have today $\rho_\varphi = 6.98 \times 10^{-121}M_{pl}^4$, $H = 1.44 \times 10^{-33}eV$, $\omega = -0.96$, and $m^2 = -\left( 2.36  \times 10^{-33}\,eV\right)^2$ and $-\left( 6.14 \times 10^{-33}eV\right)^2$.   For $\phi_{init}=0.53$ we have the current values $\rho_\varphi = 6.86 \times 10^{-121}M_{pl}^4$, $H = 1.43 \times 10^{-33}eV$ and $\omega = -0.95$ and $m^2 = \left( 2.27 \times 10^{-33}eV\right)^2$, and $- \left(4.37 \times 10^{-33}eV\right)^2$.  Quantum fluctuations, $\Delta \varphi \sim H/2\pi$, stay within the viable window close to the inflection point, right up to $H \lesssim 0.01 M_{pl}$. 
 
It is interesting to note that, independently of the initial conditions, the late time behaviour of the system far in the future, as $N \rightarrow \infty$, is:
\bea
&&\phi(N) \rightarrow \frac{1}{2a}\ln\left(\frac{36 A^2 \alpha}{5 H_0^2 \Omega_M}\right)+ \frac{3}{2\alpha} \left(N + \ln(N) \right) \nn \\
&&\rho_{\varphi} \rightarrow e^{-3N} \frac{2 H_0^2 \Omega_M}{3 N^2} \nn \\
&& \omega \rightarrow -\frac14 \,.
\eea
We have verified this both analytically and numerically.

\bigskip

We end this section with some comments on the finely-tuned initial conditions.  The required fine-tuning of initial conditions, placing the saxion close to its hilltop in order to match observations where $\omega \sim -1$, makes the scenario less compelling.  Moreover, so far, we have neglected any contributions from the matter sector to the vacuum energy, and assumed that there exists some as yet undiscovered solution to the cosmological constant problem.  However, since the scenario is based on a well-controlled landscape of supersymmetric Minkowski solutions, we should ask if there is an anthropic explanation for the initial conditions and the cosmological constant problem.

For example, notice that if the contributions from the matter sector to the vacuum energy are large and negative, $\Lambda_{M} = -M_{EW}^4$ (see \cite{Martin:2012bt} for a review), and these were the only contributions to the vacuum energy, then this would drive the Universe to recollapse on a timescale $t_\Lambda \sim 1/M_{EW} \sim 10^{-27} sec$ \cite{Edwards, Bousso:2012dk}.  However, the runaway string modulus potential \eqref{E:pot} might lift this negative vacuum energy into positive values, for some range of $\phi$.  If when starting away from the hilltop, $\phi$ runs away into large negative energy densities, then the initial condition must be finely tuned to the hilltop, to avoid a collapsing Universe.  Moreover, the hilltop must be finely tuned to a vacuum energy $\lesssim 10^{-120} M_{pl}^4$, to avoid too large an accelerated expansion, which would inhibit structure formation.  So anthropics on a supersymmetric landscape has the potential to explain fine-tuning both in the initial conditions and vacuum energy.  Possible problems, however, with this anthropic scenario are that with a fine-tuning between $\Lambda_{M}$ and $V(\phi_{max})$, the mass of the saxion will be extremely large and negative compared with $H$, and its position at the hilltop will be extremely sensitive to quantum fluctuations.  Moreover, the fine-tuning between $\Lambda_{M}$ and $V(\phi_{max})$ may be difficult to achieve, as $\Lambda_M$ is expected to vary over the cosmological history, through the EW and QCD phase transitions.

\section{Axion, Axino and the Visible Sector}
Even if it is still frozen today, the runaway saxion quintessence field is distinct from a cosmological constant as it comes with its supersymmetric partners, the axion and axino.
 
The axion remains a flat direction, until lifted by subleading non-perturbative corrections.  Assuming that subleading non-perturbative effects are sufficiently small to leave the frozen/thawing saxion undisturbed, the axion acquires an even lighter mass than the saxion, $m_\theta < m < H_0$ and similarly contributes to dark energy.

The mass acquired by the axino goes as:
\be
m_{axino} \sim 2 \phi^2 e^{K/2} D_\Phi D_\Phi W\,,
\ee
and thus behaves as dark radiation.  E.g. for the parameters given above and initial saxion values at the hilltop, the axino mass today is $m_{axino} \sim 9 \times 10^{-34}$eV.  The relic abundance would be via decay from the lightest stabilized modulus, and is model dependent.  The coexistence of saxion/axion frozen quintessence with axino dark radiation or light dark matter, might, however, provide a way to distinguish frozen quintessence from a cosmological constant\footnote{Another possible way to distinguish the frozen quintessence model from a cosmological constant is via imprints of its primordial fluctuations, although these effects are only expected to be significant in the CMB at low multipoles, where measurements are limited by cosmic variance \cite{Malquarti:2002iu, Moroi:2003pq}. Note that isocurvature contributions from saxion quintessence and its light superpartners are suppressed w.r.t.~CDM isocurvature, since their energy density is much lower than that of CDM until late times \cite{Malquarti:2002iu, Abramo:2001mv}.}.  It would also be interesting to investigate whether this or similar versions of  thawing quintessence can help to resolve the discrepancy between direct measurements of $H_0$ and those from CMB observations (see e.g.~\cite{H0, Vagnozzi:2018jhn, Lemos:2018smw}).

So far, we have focused on the moduli sector in a string scenario, with supersymmetry only very mildly broken by the runaway modulus.  Until supersymmetry is broken, the frozen quintessence vacuum energy and mass are protected by the old non-renormalisation theorems.  However, realistic string models have a visible sector where supersymmetry is broken at least at $M_{sb}\sim$TeV scales.  The effect of supersymmetry breaking in the visible sector must be sequestered from quintessence.  

This will happen if the quintessence field only couples indirectly (via virtual exchange of gravitons) to the supersymmetry breaking sector.  Assuming a tree-level decoupling between the quintessence and the visible sector, loop corrections to the mass will be of order:
\be
\Delta m^2 \sim \frac{M_{sb}^4}{M_{pl}^4} M_{sb}^2 \sim H_0^2 \,
\ee
and safe.  This requirement that the frozen quintessence be sequestered from visible matter for radiative stability, suggests that it corresponds to a modulus describing some local feature in the string compactification, with visible matter sourced elsewhere in the geometry.  E.g. blow up moduli whose flat directions are lifted by non-perturbative effects have hilltops similar to those discussed above (see Footnotes \ref{F:blowup} and \ref{F:blowupV}). In such a setup, fifth forces and time variation of fundamental constants are also evaded.

Another possible way to avoid local fifth forces and time variation of fundamental constants is if the quintessence field is frozen and universally coupled to matter.  It will be interesting to understand constraints on such models from gravitational wave experiments \cite{Yunes:2016jcc}.

\section{Discussion on dS Swampland Conjectures} \label{S:swamp}
The idea that metastable de Sitter vacua may not exist within a complete theory of quantum gravity has been gaining traction,  due in large part to the difficulties in rigorously constructing such vacua in concrete string models (see e.g. \cite{Danielsson:2018ztv, Bena:2018fqc} and references therein).   Recently this idea has been formulated as the conjectured constraints\footnote{Notice that the runaway potential discussed here seems to provide arguably the simplest counter-example to taking the original dS swampland condition \cite{Obied:2018sgi}, with only the first inequality in \eqref{E:swamp}.  For others, see \cite{Denef:2018etk, Murayama:2018lie, Choi:2018rze, Hamaguchi:2018vtv} and for earlier refinements, see \cite{Andriot:2018wzk, Andriot:2018ept}.} \cite{Obied:2018sgi,Garg:2018reu,Ooguri:2018wrx}:
\be
\frac{\sqrt{\nabla^j V \nabla_j V}}{V} \gtrsim \frac{c}{M_{pl}} \qquad \text{or} \qquad \frac{\text{min}(\nabla^i \nabla_j V)}{V} \lesssim -\frac{c'}{M_{pl}^2} \label{E:swamp}
\ee
for $c, c'$ some positive constants of order one, known as the dS Swampland Conjecture.  These
constraints rule out metastable dS vacua, but allow unstable ones.  Very recently it has been argued that they are related to the Swampland Distance Conjecture using Bousso's covariant entropy bound \cite{Ooguri:2018wrx}, see also \cite{Hebecker:2018vxz}.  See \cite{Colgain:2018wgk, Dasgupta:2018rtp, Marsh:2018kub, Wang:2018duq, Heisenberg:2018yae, Heisenberg:2018rdu, Brandenberger:2018xnf, Odintsov:2018zai, Ellis:2018xdr, Wang:2018kly, DAmico:2018mnx, Han:2018yrk} for works discussing the implications for dark energy.

The proposal for dark energy presented in Section \ref{S:hilltop} is consistent with the dS Swampland Conjecture, \eqref{E:swamp}.  Indeed, the mass-squared eigenvalues at the hilltop, for any parameters $A$ and $\alpha$, satisfy:
\be
\frac{\text{min}(\nabla^i \nabla_j V)}{V} = -2(2+ \sqrt{2}) M_{pl}^{-2} \,.
\ee
For the parameters in Section \ref{S:hilltop}, with $\phi_{init} = 0.48$, we have $\frac{\text{min}(\nabla^i \nabla_j V)}{V} = -7.3 M_{pl}^{-2}$ and for $\phi_{init} = 0.53$, we have $\frac{\text{min}(\nabla^i \nabla_j V)}{V} =  -6.2 M_{pl}^{-2}$.  Although there is a relatively large negative mass eigenvalue, we have shown explicitly that for $H \lesssim 0.01 M_{pl}$, the fields can stay within the viable window of rolling quintessence consistent with observations.  

Assuming the dS Swampland Conjecture \eqref{E:swamp} suggests that many slow-roll inflationary models are UV inconsistent.  It will therefore be interesting to investigate whether the simple runaway string moduli potentials discussed here, can also source successful inflationary models at the hilltop \cite{Boubekeur:2005zm}, consistently with the conjecture and observations.  In such models, non-standard mechanisms for reheating will be required, as for example discussed in \cite{Dimopoulos:2017tud, Dimopoulos:2018wfg}.  Another natural question is whether different classes of string moduli can source successful quintessence at the tail, thus avoiding fine-tuned initial conditions at the hilltop for quintessence, and perhaps providing a model of quintessential inflation.  Note, however, that from the discussion around \eqref{E:notail}, a slow-roll quintessence that dominates the Universe at late times requires\footnote{Note that in our conventions $\phi$ is dimensionless.} $\frac{\sqrt{3 \nabla^\phi V \nabla_\phi V}}{V} \lesssim \mathcal{O}(1)$ whereas the runaway tail is convex, so $\frac{\nabla^\phi \nabla_\phi V}{V} >0$.  Thus the dS Swampland Conjecture \eqref{E:swamp} seems to make it difficult to realise a late-time dominating slow-roll quintessence at the tail of a runaway potential, requiring in particular that the order one constant, $c$ in \eqref{E:swamp}, be less than one.

\bigskip

Of course, it is important to remember that the constraints \eqref{E:swamp} are only conjectures.  A natural question is: \emph{what might be the properties of the scalar potentials in the low energy effective field theory of consistent string compactifications that would prevent them from having metastable dS vacua?} Since most of the focus in recent discussions has been on type II constructions, we end this note by reviewing our work on dS vacua in heterotic string theory.  

In \cite{PRZ}, we computed the low energy effective field theory describing concrete heterotic orbifold constructions -- which included a scalar potential with (spontaneously broken) modular symmetry -- and searched for explicit dS vacua.  Although we identified many unstable dS vacua, we were unable to find any metastable dS vacua.  As is well known, similar experiences have been encountered in type IIA (see \cite{Roupec:2018mbn, Andriot:2018wzk} and references therein).  We have checked the dS conjecture \eqref{E:swamp} for the heterotic dS solutions in \cite{PRZ}. In particular for the double gaugino condensate example, we find $\frac{\text{min}(\nabla^i \nabla_j V)}{V}=-9.0 M_{pl}^{-2}$, while for the single gaugino condensate model $\frac{\text{min}(\nabla^i \nabla_j V)}{V}=-10.5 M_{pl}^{-2}$. Thus, these are explicit string theory dS vacua, which satisfy the dS swampland constraints, with $c' \sim 1$.   

The heterotic constructions in \cite{PRZ} had five complex moduli plus only three or four free complex parameters.  On the one hand it was easy to find dS vacua, as there were up to five non-trivial F-terms, and moreover, asking for a dS vacuum imposed only 10 real equations and one inequality on 13 or 14 degrees of freedom.  On the other hand, ensuring a metastable dS vacuum amounted to 10 real equations and 11 inequalities.  One possibility is then that we were unable to find metastable dS because there were not enough parameters in our potential.  We thus searched for metastable dS vacuum with (spontaneously broken) modular invariant scalar potentials arising from (unfortunately, an analytical study is intractable, see also \cite{Cvetic:1991qm}):
\be
K= -\ln(S+\bar{S})-3\ln(T+\bar{T}) \quad \text{and} \quad W = \frac{A_1 e^{-a_1 S} + A_2 e^{-a_2 S}}{\eta(T)^p} + \frac{B_1 e^{-b_1 S} + B_2 e^{-b_2 S}}{\eta(T)^q} + C e^{c T} \,.
\ee
Intriguingly, although there are now a sufficient number of free parameters for the four equations and five conditions for a metastable dS vacuum, we were again only able to find unstable dS, with the dS conjecture \eqref{E:swamp} again satisfied\footnote{Examples of values we obtain for $\frac{\text{min}(\nabla^i \nabla_j V)}{V}$ are $-83.5 M_{pl}^{-2}$ and $-6.8 M_{pl}^{-2}$.} for $c' \sim 1$.  However, there do exist string inspired supergravity examples of metastable dS vacua, see e.g. \cite{Covi:2008ea, Covi:2008zu, Blaback:2013qza, Kallosh:2014oja}, as well as the state of the art metastable dS string constructions discussed in \cite{Kallosh:2002gf, Cicoli:2018kdo, Kachru:2018aqn}.

\bigskip

To conclude, it remains an important problem to understand whether or not metastable dS vacua are possible within a complete theory of quantum gravity.  The refined dS Swampland conjecture -- if correct -- not only rules out metastable dS, but also makes it difficult to realise late-time dominating slow-roll quintessence at the tail of runaway scalar potentials.  However, the scenario proposed in this paper suggests that -- even if metastable dS vacua and runaway quintessence turn out to be impossible -- the unstable dS vacuum in the runaway potential of a string modulus may be able to explain the observed dark energy, with $\omega \approx -1$.

\section*{Acknowledgments}
We would like to thank Johan Bl\aa b\"ack, Kostas Dimopoulos, Fridrik Gautason, Doddy Marsh, Liam McAllister,  Viraf Metha and Radu Tatar for helpful discussions. YOT is supported by a CONACyT Mexico grant. GT and IZ are partially supported by STFC grant ST/P00055X/1. YOT thanks the Department of Physics at Swansea for hospitality.

\bibliography{refs}
\bibliographystyle{utphys}

\end{document}